\documentclass[twocolumn,showpacs,preprintnumbers]{revtex4}
\input{tcilatex}
\begin{document}

\title{The Error in the Two Envelope Paradox}
\author{Adom Giffin}
\email{physics101@gmail.com}

\begin{abstract}
The "paradox" arises in the Two Envelopes Paradox from the incorrect
formulation of the argument. \ The infomation given is misused and therefore
the results are incorrect for the question asked. \ The key is to be clear
on what question we are asking. \ We must make sure that the question that
is asked is the question that is written down.
\end{abstract}

\date{August 8, 2006}
\maketitle

\address{{\small Department of Physics, University at Albany-SUNY, }\\
{\small \ 1400 Washington Avenue, Albany, NY 12222, USA}}

\section{Introduction}

Here is the situation: There are two identical sealed envelopes, marked "A"\
and "B", that each contain a different sum of money. You are handed the
enveloped marked A. You are told that one envelope contains twice the amount
of money as the other envelope. Now you are given the choice to keep the
envelope that you have or switch it for the other envelope. In some
scenarios you are allowed to open the envelope that you have and count the
money before making a decision. So, what do you do?

Clearly being able to count the money of one of the envelopes is irrelevant
information. In other words, it does not tell you anything about the other
envelope since you were not told the sum total of the two envelopes. This
leaves us with the original situation described above.

The solution to the question has traditionally been formulated like this:
There is a one to one symmetry between the envelopes. Thus there is a 50\%
chance of getting the larger amount of money. However, if we calculate the
expectation of, let's say A, then one gets a very different result. Let us
look at the two possibilities for A. The first is that A contains twice the
amount as B ($A=2B$). The second is that A has half of what B contains ($A=%
\frac{B}{2}$). Now if we calculate the expectation value we get,%
\begin{equation}
\left\langle A\right\rangle =\frac{1}{2}(2B)+\frac{1}{2}(\frac{B}{2})=\frac{5%
}{4}B
\end{equation}%
How can the expectation value of A be greater than that of B? According to
this A should always be expected to hold more than B and one should \emph{%
always keep} the envelope. However, if A and B are switched in the
calculation the opposite occurs and one should \emph{always switch} the
envelope. Herein lies the paradox, which should you do? Keep A or trade it
for B?

\section{The error in the question}

The solution to this paradox is very simple, we are asking two different
questions and the second one is written incorrectly. The first question has
to do with which envelope we should keep. As far as \emph{this} question
goes, the \emph{only} relevant information is that the two envelopes are 
\emph{identical}. Whether one has more money than the other one is
inconsequential because you have no way of finding out what the other
envelope holds. Here is another example: Suppose someone flips a fair coin
and asks you what the outcome is going to be. You have a 50/50 chance of
being correct. Now you are told that the coin is unfair, or that the person
who flipped the coin is able to get whichever side they want. What are your
chances of being correct now? From \emph{your} perspective, it is still
50/50! The extra information is irrelevant to you because you can only pick
heads or tails. Thus, the only information that you can actually use is the
fact that the coin has two sides and as such, will not influence you about
which side to pick. So, from your perspective, it does not matter which one
you pick.

The second question asks, "What do you expect the value of the envelope you
are holding to be?" Unfortunately that is not the question that we wrote
down. The error lies in what we claim are the two possibilities. The
possibilities for envelope A are twice that of what envelope B \emph{contains%
} or one half of what B \emph{contains}. This is either two units or one
unit, since the ratio between the two is 2:1. Here we choose the first
possibility as the "two unit" possibility. The second possibility therefore
must be the "one unit" possibility which is \emph{half} of the \emph{first}
possibility. Thus the expectation should be written as follows,%
\begin{equation}
\left\langle A\right\rangle =\frac{1}{2}(2Units)+\frac{1}{2}(Units)=\frac{3}{%
2}Units
\end{equation}%
Which is what one would \emph{expect} the \emph{average} to be, exactly one
half of the total units. Expectation used in this way is, after all, an
average. And, after many trials ($n\rightarrow \infty $), one would expect
to get an average of 1.5 units. The point to understand here is that one
cannot have both 2 units \emph{and} a half of a unit \emph{simultaneously}.

\section{Conclusions}

In the end it is not the logic that usually creates a paradox, it is the
misuse or misinterpretation of it. There are several lessons to be learned
from this kind of problem. First is that one has to be very careful that the
question one is writing is indeed the question that one is asking. As has
been demonstrated, this was the case. The second lesson is understanding
when tools are useful and when they are not. For example look at the tossing
of a die. With one toss the expectation value is 3.5 units or dots. However,
this number is not even a possible outcome! But, when the die is thrown many
times the \emph{average} amount of units will indeed be 3.5. In the envelope
case illustrated above, the one half that is used for the probability in the
expectation value is really a \emph{frequency}. The \emph{probability} \emph{%
distribution} is actually flat and therefore one should have no expectation
of what the value of the envelope is at all. On the other hand when there
are many trials, as a frequency suggests, one should rightfully expect to
get the average of the envelopes which is precisely what we attained.

\begin{acknowledgments}
Thanks to Kevin Knuth for introducing me to this problem.
\end{acknowledgments}

\end{document}